\documentclass{ws-procs975x65}

\begin{document}

\title{Ramifications of the Nuclear Symmetry Energy for Neutron Stars,
Nuclei, and Heavy-Ion Collisions}

\author{A. W. Steiner}

\address{Joint Institute for Nuclear Astrophysics, National
Superconducting\\Cyclotron Laboratory, and the Department of Physics
and Astronomy,\\Michigan State University, East Lansing, MI 48824-2320}

\author{B.-A. Li}

\address{Department of Physics, Texas A\&M University-Commerce\\
Commerce, TX 75429-3011}

\author{M. Prakash}

\address{Department of Physics and Astronomy, Ohio University\\
Athens, OH 45701-2979}

\begin{abstract}
The pervasive role of the nuclear symmetry energy in establishing some
nuclear static and dynamical properties, and in governing some
attributes of neutron star properties is highlighted.
\end{abstract}


\bodymatter

\section{Introduction}

\begin{figure}[htb]
\begin{center}
\vspace*{-3.5in}
\includegraphics[scale=0.6]{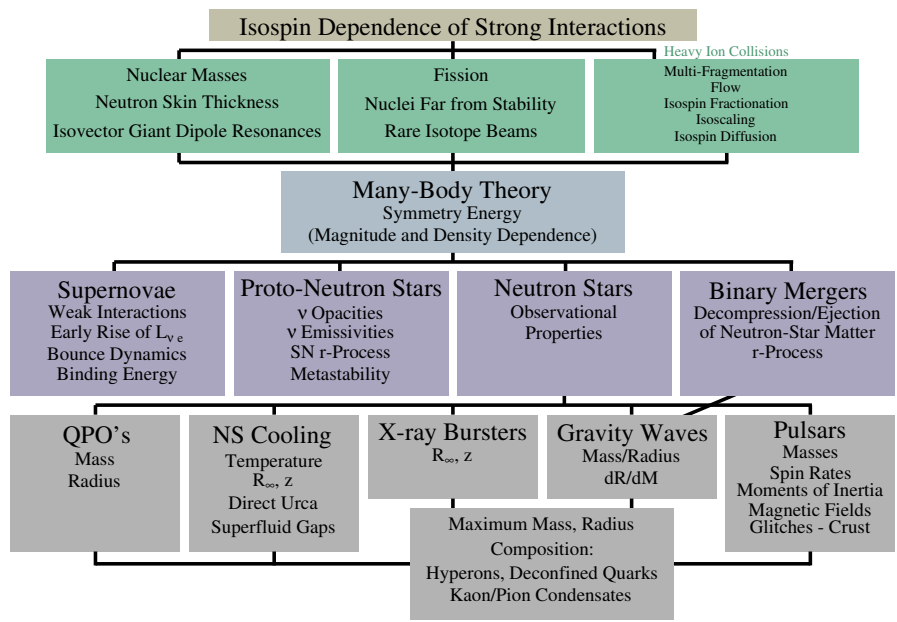}
\end{center}
\caption{The nuclear physics observables (top panels) and
astrophysical observables (lower panels) are both connected to the
nuclear symmetry energy, which is determined from nuclear many-body
physics (adapted from Ref.~\refcite{Steiner05a}).}
\label{fig:corpdiag}
\end{figure}

The nuclear symmetry energy, $E_s = \frac {1}{2} \frac {\partial^2
E}{\partial \delta^2}$, where $\delta = (n_n-n_p)/(n=n_n+n_p)$ with
$n_n$ and $n_p$ denoting the neutron and proton densities and $E
\equiv E(n,\delta)$ is the energy per particle, measures the stiffness
encountered in making a system of nucleons isospin-asymmetric.
Figure~\ref{fig:corpdiag} schematically shows how the symmetry energy
connects several nuclear and astrophysical observables.  The
difference between the energy per baryon of pure neutron matter and
that of symmetric nuclear matter (containing equal numbers of neutrons
and protons) at any particular density is largely given by the density
dependent symmetry energy. Below, we highlight some recent work that has
shed light on some of the connections between the symmetry energy and
data from terrestrial experiments and astrophysical observations.

\section{The Skin Thicknesses of Heavy Nuclei}

\begin{figure}[htb]
\begin{center}
\includegraphics[scale=0.4]{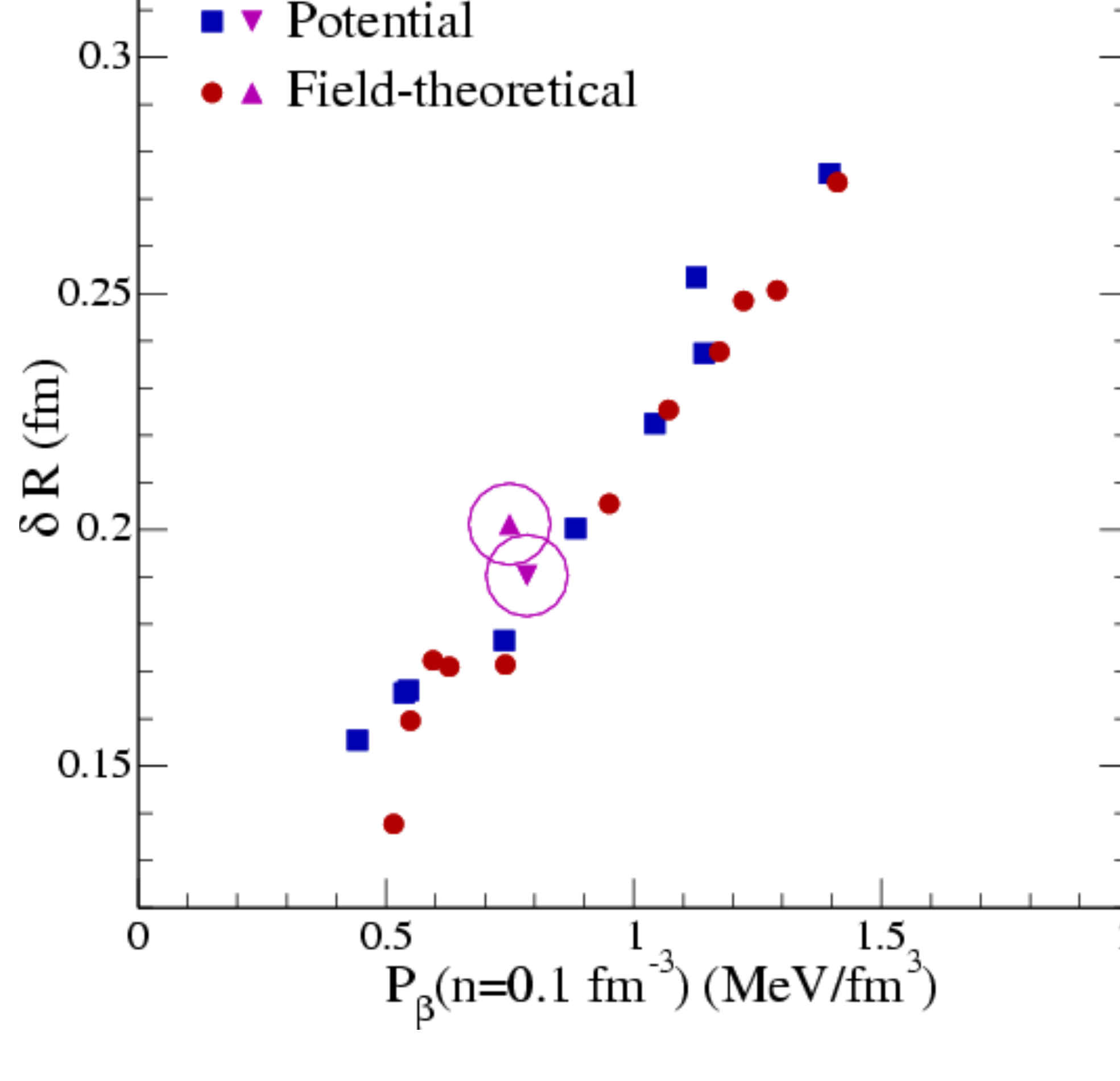}
\end{center}
\caption{The correlation between neutron skin thickness of $^{208}$Pb and 
the pressure of neutron star matter for the models described
in Ref.~\refcite{Steiner05a}.}
\label{fig:typel}
\end{figure}

Traditionally, constraints on the nuclear symmetry energy have been
derived from mass measurements of nuclei.  For neutron star physics,
the correlation that exists~\cite{Brown00,Typel01} between the neutron
skin thicknesses in heavy nuclei, $\delta R$ (the difference between
the neutron and proton root-mean-square radii), and the pressure $P$
of pure neutron matter at a density of $n \simeq 0.1$ fm$^{-3}$ is
particularly useful. Accurate measurements of $\delta R$ can
establish an empirical calibration point for the pressure of neutron
star matter at subnuclear densities. The connection between the
neutron skin thickness and the symmetry energy has been known from
Bodmer's work in the 60's~\cite{Bodmer60}. The Typel-Brown correlation
between $\delta R$ and $P(5 n_0/8)$, which is closely related to the
density derivative of the symmetry energy, demonstrates clearly the
new information that could be obtained by accurate measurements of
skin thicknesses in heavy nuclei. Figure~\ref{fig:typel} shows the
correlation between the skin thickness $\delta R$ of $^{208}$Pb and
the pressure of beta-equilibrated matter at 0.1 fm$^{-3}$ for several
potential models (based on the Skyrme interaction) and
field-theoretical models~\cite{Steiner05a}. The skin thickness of
$^{208}$Pb is scheduled to be measured at the Jefferson Lab in the
summer of 2008 in the PREX experiment~\cite{Michaels00,Horowitz01b}
and will likely provide a stringent constraint.

\section{Intermediate-Energy Heavy-Ion Collisions}

The possibility of determining the equation of state (EOS) of
nucleonic matter from heavy-ion collisions has been discussed for
almost 30 years. A number of heavy-ion collision probes of the
symmetry energy have been proposed including isospin
fractionation~\cite{Muller95,Li97}, isoscaling~\cite{Tsang01,Ono03},
neutron-proton differential collective flow~\cite{Li00}, pion
production~\cite{Li02}, isospin diffusion~\cite{Tsang04}, and
neutron-proton correlation functions~\cite{Chen03}.  Determination of
the EOS from heavy-ion data involves comparisons of transport model
simulations~\cite{Li98,Li01b,Danielewicz02} with experimental data
given an input EOS.  Transport model simulations track the evolution
of the phase space distribution function (not necessarily that of an
equilibrium distribution function at finite temperature)
at momenta that usually exceed those found in the static initial
configurations. As the driving force is the density functional
derivative of the energy density (the in-medium cross sections control
the collision integral), access to the cold EOS at high densities is
afforded.

\begin{figure}[htb]
\begin{center}
\includegraphics[scale=0.70]{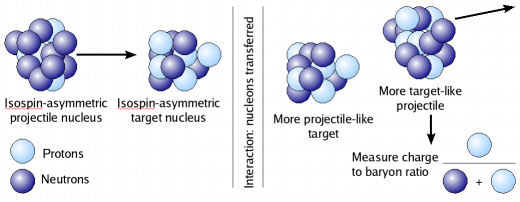}
\end{center}
\caption{Isospin diffusion process during a heavy-ion collision
(from Ref.~\refcite{Prakash06}).}
\label{fig:idiff}
\end{figure}

Isospin diffusion is caused by the exchange of neutrons and protons
between nuclei in a heavy-ion collision (Fig.~\ref{fig:idiff}). This
diffusion process, driven by the symmetry energy, moves the 
target and projectile nuclei toward isospin symmetry. To gain access to details
of the diffusion process, fragment emission during and after the
collision must be taken into account. This requirement is achieved by
considering the ratio~\cite{Rami00}
\begin{equation}
R_{\delta}=\frac{2{\delta}^{A+B}-{\delta}^{A+A}-{\delta}^{B+B}}
{{\delta}^{A+A}-{\delta}^{B+B}} \,, \label{Ri}
\end{equation}
where $A$ and $B$ denote nuclei with different isospin asymmetries and
$\delta$ is the isospin asymmetry of the projectile-like fragment.

Recently, isospin diffusion has been exploited to constrain the
symmetry energy from reactions involving $^{112}$Sn and $^{124}$Sn at
the NSCL~\cite{Tsang04,Chen05}, leading to $R_{\delta} \sim 0.46$.  In
conjunction with an isospin- and momentum-dependent transport model,
IBUU04~\cite{Li04,Li04b}, the NSCL data on isospin diffusion can be
used to constrain the symmetry energy. Writing the symmetry energy in
terms of a kinetic part and a potential part as
\begin{equation}
E_{\mathrm{s}}(n) = S_{v} (n/n_0)^{\gamma} \,,
\end{equation}
where $S_{v}$ is the symmetry energy at the nuclear equilibrium
density, $n_0=0.16~{\rm fm}^{-3}$, the constraint $0.69<\gamma<1.05$
has been found~\cite{Chen05,Li05c}. This constraint is consistent with
the symmetry energy inherent in the EOS (computed using Monte Carlo
simulations with input two- and three-body interactions which are
matched to nucleon-nucleon scattering phase shifts and the energy
levels of light nuclei) of Akmal, et al.~\cite{Akmal98} (APR). This
constraint also rules out models with values of $\gamma>1.05$ found in
some field-theoretical models.

\begin{figure}[htb]
\begin{center}
\includegraphics[scale=0.5]{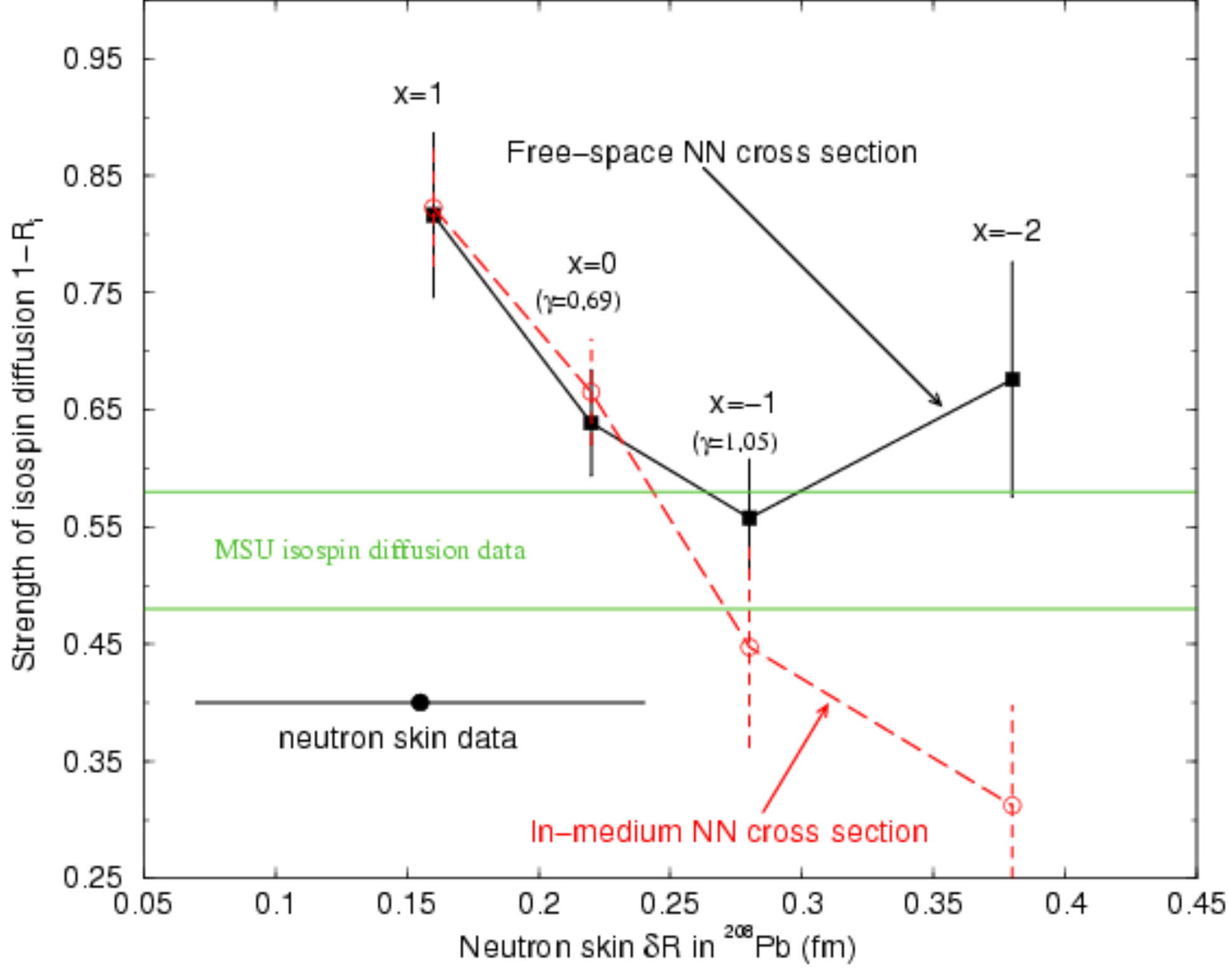}
\end{center}
\caption{The correlation between the neutron skin thickness and the
amount of isospin diffusion in collisions between Sn isotopes as
measured at the NSCL (from Ref.~\refcite{Li05c}). }
\label{fig:nskin}
\end{figure}

Because intermediate-energy heavy-ion collisions provide a constraint
on the symmetry energy at the same densities as would be probed by a
measurement of the neutron skin thickness of $^{208}$Pb, the NSCL data
also provides a restrictive range for its neutron skin thickness. This
connection was used in Refs.~\refcite{Steiner05b} and \refcite{Li05c}
to show that the neutron skin thickness of $^{208}$Pb should be at
least greater than 0.15 fm in order to be consistent with the NSCL
data, with values between 0.23 fm and 0.27 fm favored by the transport
model simulations (Fig.~\ref{fig:nskin}).

\section{Neutron Star Radii}

Neutron star radii tend to probe the density dependence of the
symmetry energy around the nuclear equilibrium density, $n_0=0.16~{\rm
fm}^{-3}$.  Lattimer and Prakash \cite{Lattimer01} found that the
radius $R$ of a neutron star exhibits the power law correlation:
\begin{equation}
R \simeq C(n,M)~[P(n)]^{0.23-0.26}\,,
\label{correl}
\end{equation}
where $P(n)$ is the total pressure inclusive of leptonic contributions
evaluated at a density $n$ in the range $n_0$ to 2$n_0$, and $C(n,M)$
is a number that depends on the density $n$ at which the pressure is
evaluated and on the stellar mass $M$. The left panel
in Fig.~5 shows
this correlation as $RP^{-\alpha}$ versus $R$ for stars of mass
$1.4~{\rm M}_\odot$. Neutron star radius measurements, especially
those with uncertainties less than about 0.5 km, constrain the
symmetry energy above the nuclear equilibrium density. These
constraints will be much improved when simultaneous mass and radius
measurements of the same object become available.

\begin{figure}[htb]
\begin{center}
\vspace*{-1.0in}
\parbox{2.6in}{\epsfig{figure=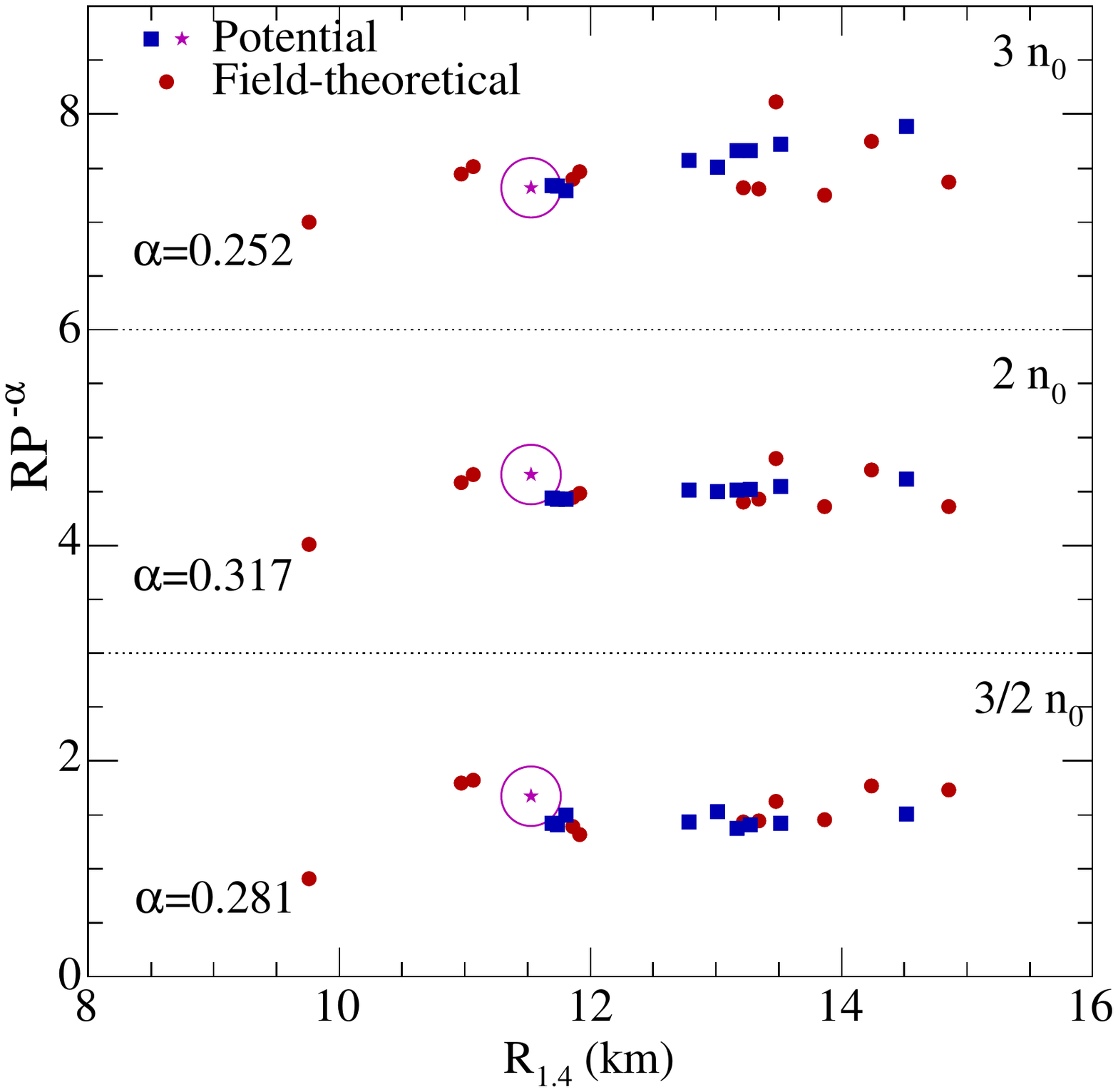,width=3.0in}}
\parbox{2.1in}{\epsfig{figure=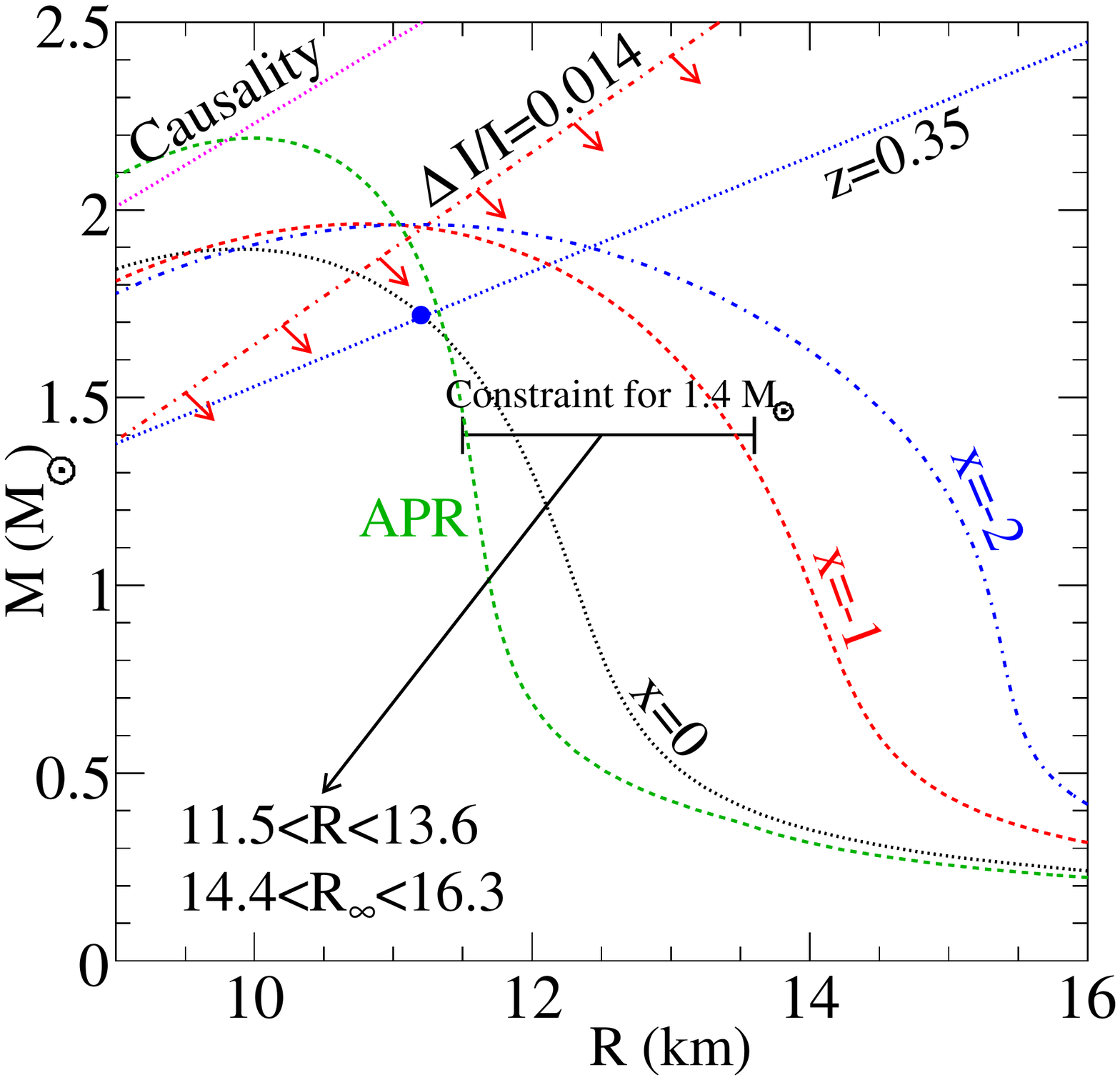,width=2.4in}}
\caption{{\it (Left)} The value $ RP^{-\alpha} $ for several
field-theoretical and potential models considered in
Ref.~\refcite{Steiner05a}. The values of $\alpha$ correspond to the
density indicated in the upper right corner of each block. {\it
(Right)} The constraint on the radius of 1.4 solar mass neutron stars
as determined from isospin diffusion in heavy-ion collisions (from
Ref.~\refcite{Li05}).}
\end{center}
\label{fig:radfig}
\end{figure}

If there is no phase transition between $n_0$ and a few times $n_0$,
the range in which neutron star radii are determined mainly by the
symmetry energy, results from the isospin diffusion data at the NSCL
can be used to constrain neutron star radii~\cite{Li05}.  As only EOSs
with symmetry energies between $x=0$ and $x=-1$ (where $x$ is a
parameter designed to vary the density dependence of the symmetry
energy without modifying the magnitude of the symmetry energy at $n_0$
or the isospin-symmetric part of the EOS) are consistent with the
isospin diffusion data, this range of $x$ values is representative of
the possible variation in neutron star structure that is consistent
with terrestrial data. Neutron star radii, while being strong
functions of the symmetry energy, are also affected by contributions
from the isospin-symmetric part of the EOS, especially at high
densities.  About 5\% difference is representative of the radius
uncertainty stemming from the symmetric part of the EOS.  The
conclusion of Ref.~\refcite{Li05} is that only radii between 11.5 and
13.6 km (or radiation radii between 14.4 and 16.3 km) are consistent
with the $x=0$ and $x=-1$ EOSs (see the right panel in
Fig.~5).

\section{The Direct Urca Process}
\label{DUrca}

The long-term cooling of a neutron star is chiefly determined by its
composition. Beta equilibrium and charge neutrality determine the
proton fraction in neutron-star matter and thus the critical density
for the onset of the direct Urca processes $n \rightarrow p + e + \bar\nu
$ and $e + p \rightarrow n + \nu$, which cool the star more rapidly
than the modified Urca processes in which an additional nucleon is
present~\cite{Lattimer91b}.  The direct Urca processes, however,
require a sufficient amount of protons in matter (of order 10-14\%).

\begin{figure}[htb]
\vspace*{-1.0in}
\begin{center}
\includegraphics[scale=0.37]{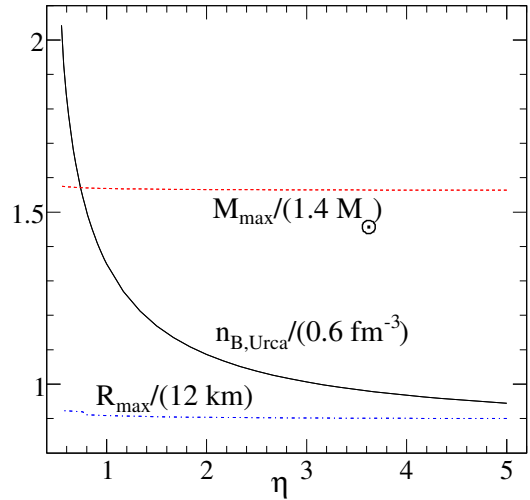}
\end{center}
\caption{The maximum mass, the radius of the maximum mass star, and
the critical density for the direct Urca process as a function of
$\eta$, which describes the strength of quartic terms in the symmetry
energy (from Ref.~\refcite{Steiner06}). }
\label{fig:quart}
\end{figure}

Larger symmetry energies induce larger proton fractions (with matter
being closer to isospin-symmetric) and smaller critical densities for
the onset of the direct Urca processes.  However, higher than
quadratic terms in the energy $E(n,\delta)$ of isospin asymmetric
matter can have an important role to play~\cite{Muether88}.  Recently,
Steiner~\cite{Steiner06} has shown that quartic terms in $E(n,\delta)$
play an important role in determining the critical density for the
direct Urca process. Such terms can be easily generated within the
context of field-theoretical models.  This is demonstrated in
Fig.~\ref{fig:quart} for the Akmal, et al., (1998) EOS, for which the
relative size of the quartic and quadratic terms is parametrized by
$\eta$ as described in Ref.~\refcite{Steiner06}. The mass and radius
are virtually unchanged, whereas the threshold density for the direct
Urca process changes by more than a factor of two.

Gusakov, et al.,~\cite{Gusakov05} have investigated the cooling
of neutron stars using the EOS of APR~\cite{Akmal98}.
They found that, because of the direct Urca process, stars with masses
larger than about 1.7 $\mathrm{M}_{\odot}$ cool so rapidly
as to be cooler than nearly all of the observed neutron stars. Our
work offers a possible resolution: quartic terms can play a role at
high density to turn off the direct Urca process thus making the
computed cooling curves match the comparatively warm neutron stars.

\section{Outlook}

In addition to the role of the symmetry energy in the few areas
highlighted here, its importance in controlling the cooling times of
transient x-ray bursters, seismic activity of neutron star surfaces,
ejection of baryons during binary mergers, etc., is only beginning to be
appreciated.  The PREX experiment, heavy-ion 
experiments, neutron star mass, radius and surface temperature
measurements, observations of transient x-ray bursters etc., all hold
keys to pin down the magnitude and density dependence of the
symmetry energy in addition to delineating its pervasive role.

\section{Acknowledgments}

Research support for A.~W.~S. by the Joint Institute for Nuclear Astrophysics
under the NSF-PFC grant PHY~02-16783, for B-A.~L. in parts
by the NSF under Grant No. PHY0652548 and the Research Corporation
under Award No.~7123, and for M.~P. by the DOE under
Grant. DE-FG02-93ER40756 is gratefully acknowledged. 

\bibliographystyle{ws-procs975x65}
\bibliography{proc3}

\end{document}